\documentclass[default,iicol,pdflatex,sn-mathphys]{sn-jnl}

\usepackage{graphicx}

\usepackage{dcolumn}
\usepackage{bm}
\usepackage[utf8]{inputenc}
\usepackage[T1]{fontenc}
\usepackage{mathptmx}
\usepackage[detect-weight=true, detect-family=true]{siunitx}
\usepackage{numprint}
\DeclareSIUnit\torr{Torr}
\usepackage{caption}
\usepackage{commath}

\newcommand{\mps}[0]{\meter\per\second}
\newcommand{\mpss}[0]{\meter\per\second\squared}
\newcommand{\nm}[0]{\nano\meter}
\newcommand{\ns}[0]{\nano\second}

\newcommand{\cm}[0]{\centi\meter}

\jyear{2022}%

\theoremstyle{thmstyleone}%
\theoremstyle{thmstyletwo}%

\theoremstyle{thmstylethree}%

\begin{document}

\title[Optical deceleration of atomic hydrogen]{Optical deceleration of atomic hydrogen}


\author*[1]{\fnm{Samuel F.} \sur{Cooper}}\email{samuel.cooper@rams.colostate.edu}

\author[1]{\fnm{Cory} \sur{Rasor}}\email{crasor@rams.colostate.edu}

\author[1]{\fnm{Ryan G.} \sur{Bullis}}\email{ryan.bullis@colostate.edu}

\author[1]{\fnm{Adam D.} \sur{Brandt}}\email{adam.brandt@nist.gov}

\author[1]{\fnm{Dylan C.} \sur{Yost}}\email{dyost@colostate.edu}

\affil*[1]{\orgdiv{Department of Physics}, \orgname{Colorado State University}, \orgaddress{\street{1875 Campus Delivery}, \city{Fort Collins}, \postcode{80523}, \state{Colorado}, \country{USA}}}







\maketitle

\fontfamily{lmss}\selectfont
\textbf{
High-precision hydrogen spectroscopy is an active field which helps to determine the Rydberg constant and proton charge radius \cite{tiesinga_codata_2021}, tests bound-state QED \cite{karshenboim_precision_2005}, and can search for Beyond Standard Model (BSM) Physics \cite{karshenboim_precision_2010,yang_probe_2020,jones_probing_2020}.  Additionally, with recent demonstrations of anti-hydrogen trapping and spectroscopy, a new line of investigation is possible whereby hydrogen can be compared to its antimatter counterpart \cite{bluhm_mathitcpt_1999,atrap_collaboration_first_2004,andresen_trapped_2010,atrap_collaboration_trapped_2012,kostelecky_lorentz_2015,malbrunot_asacusa_2018,ahmadi_characterization_2018,the_alpha_collaboration_investigation_2020}. The next generation of precision hydrogen spectroscopy will likely require additional motional control of the atomic sample \cite{jones_probing_2020} -- similar to what is possible with heavier elements \cite{metcalf_laser_2007}. Unfortunately, laser cooling – one of the cornerstones of modern precision atomic physics – is difficult in hydrogen due to the vacuum ultraviolet radiation required \cite{hijmans_optical_1989,eikema_continuous_2001,michan_development_2014,gabrielse_lyman-_2018}.  Here, we sidestep the challenges inherent in laser cooling and demonstrate a technique whereby we load metastable atoms from a cryogenic beam into a moving optical lattice, decelerate the lattice, and observe a commensurate deceleration of the atoms.  Since the optical lattice is governed by standard optoelectronics, this technique represents a robust platform for the motional control of hydrogen.  Our technique could enable greater precision in hydrogen spectroscopy and be transferred to exotic simple atoms such as antihydrogen \cite{malbrunot_asacusa_2018,malbrunot_hydrogen_2019}, and muonium \cite{janka_intense_2020,crivelli_mu-mass_2018,ohayon_precision_2022,delaunay_towards_2021}.}

\fontfamily{ptm}\selectfont

\begin{figure}[t!]
    \centering
    \includegraphics[width=75mm]{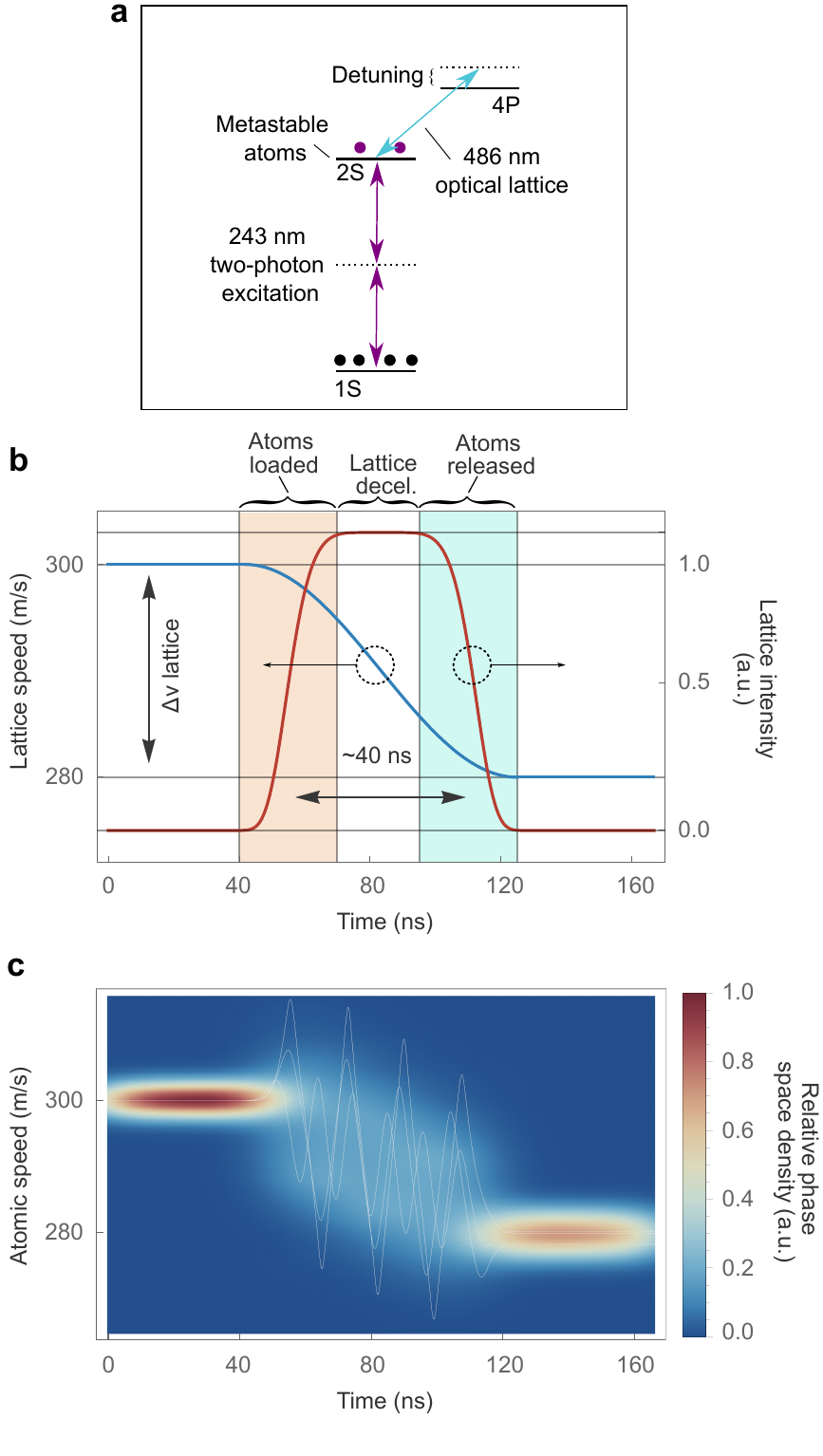}
    \caption{\textbf{Method for the optical deceleration of atomic hydrogen.} \textbf{a}. Simplified energy level diagram showing relevant levels. After excitation to the 2S state using two photon excitation, an optical lattice is produced using light nearly resonant with the 2S-4P transition. \textbf{b}. Timing of our slowing method.  The lattice intensity is shown in red, whereas the lattice velocity is shown in blue. The red shaded region shows the loading phase where atoms comoving with the lattice at \SI{300}{\mps} are captured.  The lattice is then decelerated to slow the trapped atoms before the lattice intensity is ramped down in the blue shaded region. \textbf{c}. Density plot of simulated atomic trajectories over the course of a deceleration pulse for atoms with an initial velocity near \SI{300}{\mps}. The colors represent the atomic densities in phase space. As the lattice decelerates from \SI{300}{\mps} to \SI{280}{\mps} the atoms oscillate within the wells of the optical lattice before their release. Three example atomic trajectories (white tracks) overlay the density plot.
    }
    \label{fig:intro}
\end{figure}

\begin{figure}[thb]
    \centering
    \includegraphics[width=75mm]{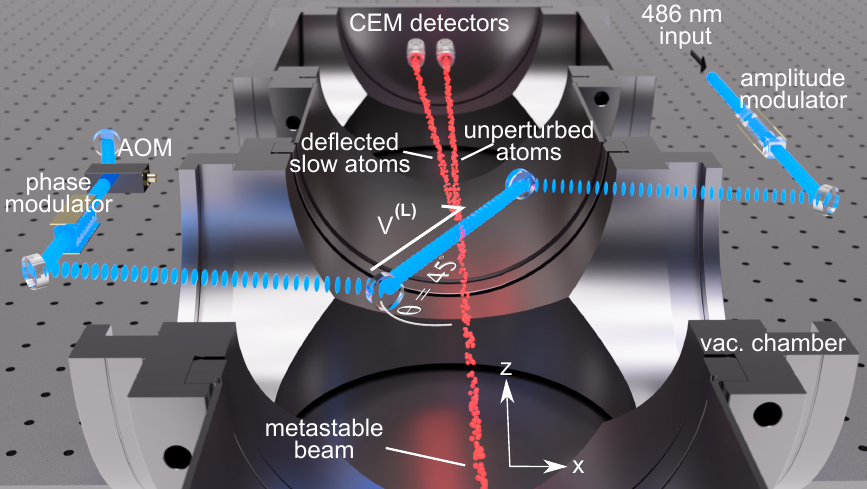}
    \caption{\textbf{Experimental setup.} Balmer-$\beta$ light (\SI{486}{\nm}) first passes through an amplitude modulator which controls the lattice intensity. The light then passes through the interaction region before being retroreflected. The retroreflected beam is double-passed through an AOM to establish a central lattice velocity, and a phase modulator to provide the lattice acceleration. The lattice is placed at a \SI{45}{\degree} angle to the atomic beam  so that the lattice simultaneously slows atoms along the beam axis (the z direction) and deflects them in the x direction. Two detectors count metastable atoms. One is placed on axis to provide normalization, and a second is placed at $x=$\SI{-2}{\centi\meter}, \SI{+27}{\cm} after the deceleration region.  With this, atoms decelerated by \SI{20}{\mps} will be incident on the off-axis detector. 
    }
    \label{fig:apparatus}
\end{figure}

The study of hydrogen has played an important role in modern physics and was instrumental in the development of quantum mechanics, the Dirac equation, and quantum electrodynamics (QED) \cite{lamb_fine_1947}. In addition to this illustrious history,  hydrogen spectroscopy remains a dynamic field where measurements have reached a very high level of precision.  As an example, the 1S-2S transition has been measured with an impressive fractional uncertainty of only 4.2 $\times$ 10$^{-15}$ \cite{parthey_improved_2011}. 
However, most precision hydrogen spectroscopy is conducted with atomic beams with temperatures between 4.5 K and room temperature \cite{parthey_improved_2011,beyer_rydberg_2017,fleurbaey_new_2018,bezginov_measurement_2019,grinin_two-photon_2020,brandt_measurement_2022} -- at least six orders of magnitude larger than what is routinely achieved in heavier atomic species through laser cooling \cite{metcalf_laser_2007}. This relatively high temperature often limits the measurement uncertainty, which has motivated continued research into new methods to laser cool and decelerate hydrogen beams. The most conceptually simple method to laser cool hydrogen is through the 1S-2P cycling transition using Lyman-$\alpha$ radiation at \SI{121.6}{\nm}.  However, Lyman-$\alpha$ radiation is very challenging to produce and manipulate \cite{hijmans_optical_1989,eikema_continuous_2001,michan_development_2014,gabrielse_lyman-_2018}. Therefore, while laser cooling of hydrogen and antihydrogen has been demonstrated  \cite{hijmans_optical_1989,baker_laser_2021}, it is still far from routine. This has motivated other methods to achieve ultracold atomic hydrogen such as pulsed two-photon Doppler cooling \cite{kielpinski_laser_2006,wu_pulsed_2011} or photodissociation of trapped hydride molecules \cite{mcnally_optical_2020,vazquez-carson_direct_nodate}. In addition, there have been methods proposed \cite{raizen_comprehensive_2009,narevicius_stopping_2008} and demonstrated \cite{hogan_magnetic_2008} to decelerate supersonic hydrogen beams using pulsed magnetic fields. 

An intriguing option for the deceleration of atomic or molecular beams is to use the optical dipole force \cite{dahan_bloch_1996,fulton_optical_2004, fulton_controlling_2006}. However, to the best of our knowledge this has been unexplored with atomic hydrogen. This is likely because large dipole forces are typically only achievable with radiation which is nearly resonant to an electronic transition and, for ground-state hydrogen, these transitions are in the vacuum-ultraviolet spectral region, which would introduce challenges similar to those for laser cooling. 

Here, we circumvent these challenges by first exciting our atomic beam to the metastable 2S state (lifetime $\sim$ 122 ms).  With that, there are several electronic transitions which can be accessed with visible lasers producing large optical forces.  For the experiments described in this letter we use high-power ($\sim$\SI{1}{\watt}) continuous-wave (cw) laser radiation at 486 nm which is nearly resonant with the 2S-4P transition. Our deceleration method uses a moving optical lattice in the path of our metastable hydrogen atomic beam. The slowing consists of three steps which are shown in Fig.\ref{fig:intro}. First, the lattice intensity is quickly turned on trapping a fraction of the atoms. For this step, the speed of the lattice is set near the peak speed of the atomic beam, and the fraction of the velocity distribution trapped is determined by the depth of the lattice wells. The lattice is then decelerated, which produces a corresponding deceleration of the trapped atoms. Finally, the lattice is quickly turned off, leaving the atoms with a reduced speed. For our demonstration, atoms moving at speeds of roughly $300$ m/s $ \pm 15$ m/s are trapped in the lattice wells and the atoms are then decelerated by about 20 m/s in a \SI{40}{\ns} slowing cycle. Atoms outside of this 30 m/s velocity range will typically travel between many lattice sites and are usually left with a velocity relatively unaffected by the slowing procedure.
    
The design of our slowing process was determined using simple atomic physics arguments. For these estimates, we ignore fine and hyperfine structure and consider only the 2S and 4P levels. For an optical lattice which is far-detuned from the 2S-4P resonance, the maximum lattice potential is given by 
\begin{equation}
    U_{max}=\hbar \frac{\mid\Omega_{max}\mid^2}{4 \delta},
\end{equation}
where $\Omega_{max}$ is the maximum Rabi frequency, and $\delta$ is the detuning from the resonance in angular frequency units.  From this, it is clear that a larger well-depth can be obtained by increasing $\Omega_{max}$, or decreasing $\delta$. In general, it is desirable to maximize $U_{max}$ to slow a larger fraction of the thermal atomic beam's velocity distribution.  However, it is also important to consider quenching of the metastable 2S state via spontaneous emission. The dipole force originates from a small mixing of the 2S and 4P states, and the 4P state can decay to the ground 1S state. Once an atom is in the ground state, it ceases to be affected by the dipole forces from the optical lattice. Therefore, an optimized slowing process should be designed to minimize quenching of the 2S state while maximizing the optical lattice potential well depth. The maximum quench rate is given by 
\begin{equation}
    \Gamma_{quench}=\frac{\mid\Omega_{max}\mid^2}{4 \delta^2} \gamma_{4P},
\end{equation}
where $\gamma_{4P} \approx \SI{8.1e7}{\per\second}$ is the decay rate from the 4P state.  From these considerations, it is clear that the performance of the slower will generally be improved with a larger laser intensity, which increases $\Omega_{max}$, since $\delta$ cannot be arbitrarily reduced without unacceptable loss of metastable 2S population.  In addition, there is an advantage to a fast deceleration process to limit quenching of the metastable atoms.

\begin{figure}[t!]
    \centering
    \includegraphics[width=75mm]{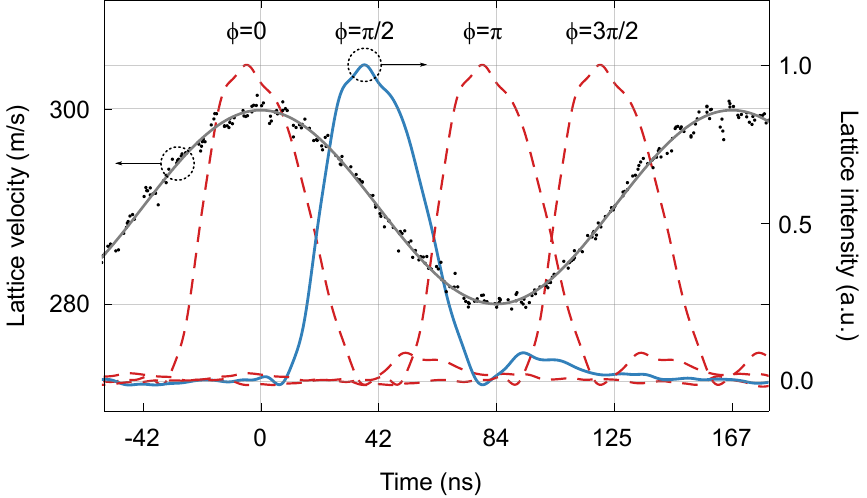}
    \caption{\textbf{Experimental control of the optical lattice velocity.} The measured sinusoidal modulation of the lattice velocity in the direction of the atomic beam is shown in black ($a=$\SI{4e8}{\mpss}). The relative phase between the lattice velocity modulation and the lattice amplitude is given by $\phi$. The phase $\phi=\pi/2$ (solid blue) is the standard condition for a deceleration. We can also vary the phase to other values such as $\phi=0$, $\pi$, and $3\pi/2$ (dashed red) to demonstrate our control over the slowing process. 
    }
    \label{fig:MCphase}
\end{figure}

Our experimental demonstration of this slowing method begins with a cryogenic atomic beam (\SI{5}{\kelvin}), which has a peak velocity near \SI{470}{\mps} \cite{cooper_cryogenic_2020}.  Laser radiation at \SI{243}{\nm} crosses the atomic beam to excite the 1S-2S two-photon transition producing a metastable beam of 2S hydrogen \cite{cooper_cavity-enhanced_2018,cooper_cryogenic_2020,brandt_measurement_2022}.  The 2S hydrogen then travels for approximately \SI{15}{\cm} before arriving at the deceleration region.  Here, the atomic beam crosses an optical lattice produced with $\sim$ \SI{1}{\watt} of retro-reflected laser radiation at \SI{486}{\nm}. As shown in Fig. \ref{fig:apparatus}, the optical lattice is oriented at a \SI{45}{\degree} angle with respect to the $z$-axis (i.e. the initial atomic beam propagation direction). With this, the atoms receive a horizontal deflection in the $x$-direction as they are being slowed. This allows for a convenient method to detect the slowed fraction of the atomic beam with reduced background. Metastable hydrogen is readily detected with a simple channel electron multiplier (CEM) \cite{brandt_measurement_2022}.  We use two such detectors -- one in the direction of the initial atomic beam, and a second off-axis which is set to detect the slowed and deflected atoms. 
An acousto-optic modulator and electro-optic phase modulator in the retro-reflected path allows for precise control of the velocity and acceleration of the optical lattice. For the demonstration here, the 486 nm radiation is shifted by a variable frequency, $\Delta f$, of between \SI{815}{\mega\hertz} and \SI{872}{\mega\hertz}.  This produces a corresponding lattice well speed, given by $\mid~v^{(L)}~\mid =\Delta f/2 \times 486$ nm, of between \SI{198}{\mps} and \SI{212}{\mps}.  However, since the lattice is oriented at 45$^\circ$ with respect to the $z$-axis (the atomic beam direction), the lattice velocity with respect to that axis is more relevant. This is given by $v_z^{(L)}=\sqrt{2} \, \mid {v^{(L)}}\mid $, which varies from \SI{280}{\meter\per\second} to  \SI{300}{\meter\per\second}. 

The 486 nm radiation is focused to a spot size of \SI{80}{\micro\meter} and is tuned to within about \SI{1.2}{\giga\hertz} of the 2S-4P transition.  With this lattice intensity and detuning, atoms which are traveling within $\pm$\SI{15}{\meter\per\second} of the initial lattice speed can be trapped within the lattice wells. As shown in Fig. \ref{fig:apparatus}, we can precisely control the depth of the optical lattice or turn it off completely using an electro-optic amplitude modulator on the \SI{486}{\nm} radiation before it reaches the slowing region.  Since it takes roughly \SI{1}{\micro\second} for the atomic beam to traverse the slowing region, the slowing procedure can be repeated at rates up to \SI{1}{\mega\hertz}.  With this, a given decelerated atom is allowed to exit the interaction region before the next deceleration sequence is applied. 
 
 One of the attractive features of this slowing technique is that the acceleration of the lattice and its intensity are precisely controlled with standard optoelectronic devices and RF electronics. To demonstrate and characterize the control of this technique we applied the lattice pulse with a variable phase, $\phi$, with respect to a sinusoidally varying lattice velocity.  This modulation had a \SI{167}{\nano\second} period and an amplitude of \SI{10}{\mps} centered at a mean velocity of \SI{290}{\mps}. Depending on this relative phase, the lattice can be accelerating, decelerating, or held stationary while it is pulsed on. The experimentally measured lattice amplitude and lattice velocity are shown in Fig. \ref{fig:MCphase} for several values of $\phi$. We then measured the number of counts on the off axis atom detector while varying this phase, holding all other parameters constant. When the pulse has a phase of $\phi=\pi/2$ the lattice is decelerating, which reproduces the desired operation of our cooling scheme as depicted in Fig. \ref{fig:intro}.  
 
\begin{figure}[t!]
   \centering
   \includegraphics[width=75mm]{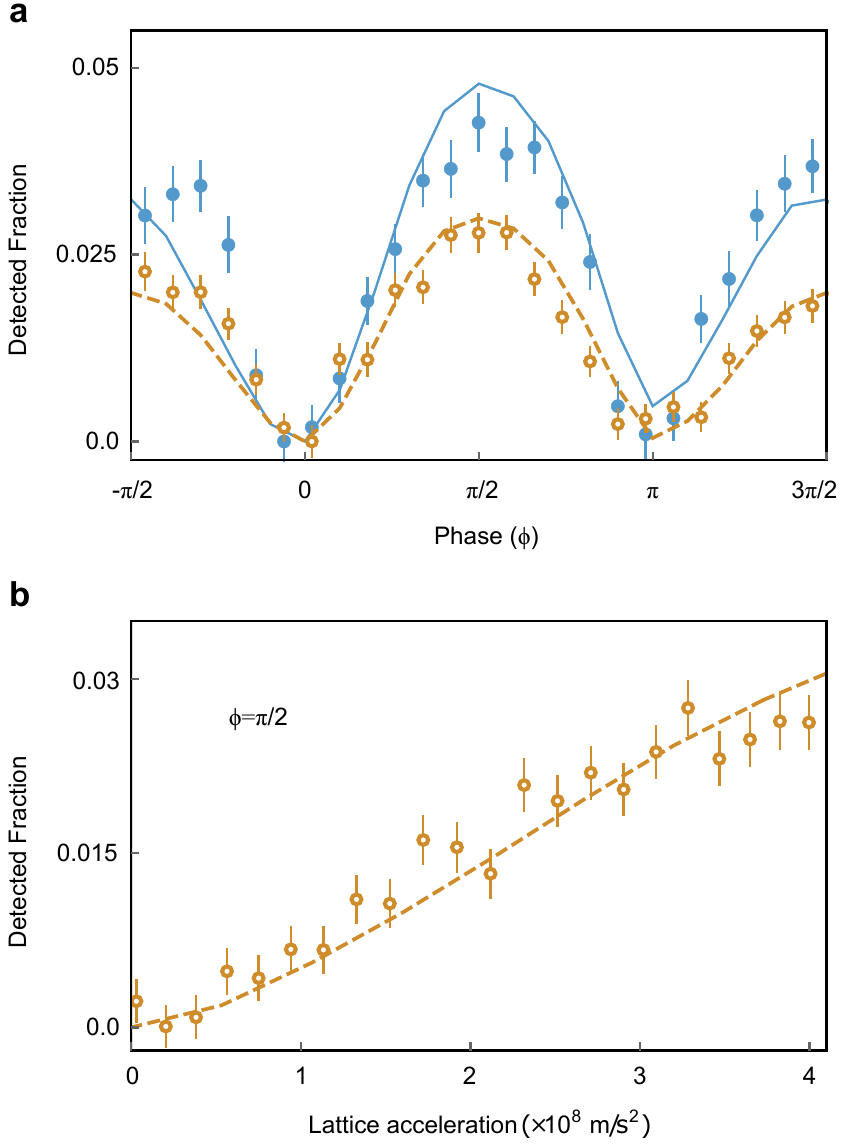}
   \caption{\textbf{Observation of slowed-atom signal as function of lattice parameters.} \textbf{a}, Measured fraction of atomic beam deflected into the off-axis detector as a function of $\phi$. Blue solid points were taken at a detuning of $\delta=2 \pi \times$\SI{1.2e9}{\per\second}, a central lattice velocity of $v_z^{(L)}=$\SI{274}{\mps}, and a lattice acceleration of 4 $\times \, 10^8$ ms$^{-2}$.  The solid blue line shows a simulation for the same experimental conditions. The yellow hollow data points were taken at a detuning of $\delta=2 \pi \times$\SI{1.5e9}{\per\second}, with a central lattice velocity $v_z^{(L)}=$\SI{307}{\mps}, and a lattice acceleration of \SI{4e8}{\mpss}.  The dashed yellow line shows a simulation for these same experimental conditions. \textbf{b}, The off-axis signal as a function of the peak lattice acceleration.  Data and simulation are for the same conditions as the yellow hollow points in \textbf{a}. In both plots, the detected fraction is relative to the number of atoms within 30 m/s velocity near the central lattice velocity -- i.e. the number of trappable atoms. This fraction was determined with the velocity distribution measured in \cite{cooper_cryogenic_2020}. Each data point is measured with 5 seconds of integration time and the error bars are calculated from the shot noise. All simulations use only experimental conditions with no free fit parameters. 
    }
    \label{fig:results}
\end{figure}

Figure \ref{fig:results} shows the slowed atom signal in the off-axis detector as a function of $\phi$, where each data point corresponds to an integration time of 5 seconds.  As expected, we observe a peak in the slowed-atom signal when the lattice is pulsed on as it is decelerating ($\phi=\pi/2$). With this phase, the signal measured by the off-axis channeltron is regularly $\sim$100 counts per second which corresponds to roughly 5\% of the slowable atoms (i.e. within 30 m/s near the central lattice velocity) in the measured velocity distribution \cite{cooper_cryogenic_2020}.  This fraction is experimentally reasonable since our metastable atomic beam is roughly 350 $\mu$m in height whereas whereas the optical lattice is only roughly 80 $\mu$m in height and the repetition rate of the slowing process is not fast enough to ensure that all atoms in our continuous beam experience the peak lattice field. This is confirmed with our experimental simulations discussed later. When varying $\phi$, we see a minimum in the slowed-atom signal when the lattice is pulsed on without an acceleration ($\phi=0, \pi$). Interestingly, we also see a significant slowed-atom signal when $\phi=3\pi/2$, which corresponds to the condition where the lattice has a positive acceleration in the direction of the atomic beam velocity when it is pulsed on.  

While initially surprising, this behavior can be understood intuitively. For an atom to be localized at a single lattice site, an atom must be within \SI{30}{\mps} of the initial lattice velocity. An atom which is slightly faster than this will travel between several lattice sites.  However, if the lattice is accelerating in the same direction as the atomic velocity, then at some point in the interaction, the atom could be traveling with a peak speed only slightly faster than the lattice and will be reflected from the side of the nearest lattice well.  This results in an impulse to the atom even when $\phi=3\pi/2$.  This behavior is similar to the slowing observed when atoms are reflected from a moving surface \cite{narevicius_coherent_2007}.

To gain insight into this behavior, we performed a Monte Carlo (MC) simulation of the slowing process. For this, the atomic motion in the presence of the time-dependent lattice potential, $U(\vec{r},t)$, was treated classically.  In addition, our simulation calculated the probability that the 2S population would be lost due to spontaneous emission through the 4P-1S transition. 
We integrated our simulation over the initial atomic positions and velocities present in our hydrogen beam and the results of the simulation are shown in Fig. \ref{fig:results} overlapped with our experimental results.  As can be seen in the figure, we achieve excellent agreement when our simulation is seeded with our experimental parameters.

We believe the slowing method demonstrated in this letter is flexible and could easily follow the geometry in \cite{fulton_controlling_2006} to slow without deflection, or even placed at an angle to the vertical plane to deflect atoms into a fountain style trajectory to perform Ramsey spectroscopy \cite{huber_two-photon_1998}. Here we demonstrate a reduction in the kinetic energy of slowed atoms by roughly 12\%.  However, we believe bringing the atoms to a near standstill is possible. Given the simple atomic physics arguments above, it is clear that a larger deceleration without additional quenching, or the slowing of a larger portion of our velocity distribution would be aided by additional lattice intensity. Previously, we have demonstrated \SI{4.2}{\watt} of optical power at \SI{486}{\nm} enabled with a Yb fiber amplifier \cite{burkley_highly_2019}, which could greatly improve performance.  As discussed in the Methods section, here we used a much simpler \SI{486}{\nm} laser system for the optical lattice.  
In addition, it is notable that by using a CW laser along with an intensity modulator with a low duty cycle, 95\% of the optical power is not used for slowing.  Therefore, it is interesting to consider the use of a specially designed Q-switched laser to efficiently use the optical power available. However, the use of a CW laser as demonstrated here is more flexible since we have full control over the lattice intensity and pulse duration with relatively simple electronics. 

As discussed in the introduction to this letter, slow atomic samples of hydrogen could improve the precision of hydrogen spectroscopy, and in turn, yield increased sensitivity for tests of fundamental physics. 
Since the highest precision spectroscopy measurements of hydrogen have been have been performed with atomic beams with temperatures $>$\SI{4.5}{\kelvin}, this work could drastically improve the systematic uncertainties related to atomic velocity in these types of measurements. In addition, this slowing method could be used to load a magnetic atomic trap \cite{hogan_magnetic_2008} or an optical dipole trap trap \cite{kawasaki_magic_2015}.  
Finally, it is interesting to note that exotic atomic systems such as anti-hydrogen and muonium have indentical, or near-identical, energy level structure as hydrogen.  Therefore, the challenges in the optical  manipulation of the motion of these atoms is similar to hydrogen, and we believe that our technique could be adapted to such experiments.  A comparison of the Rydberg constant extracted in these different systems are of very valuable \cite{jones_probing_2020} since they can provide proton structure-free tests of QED and a high precision measurement of the muon mass in muonium \cite{crivelli_mu-mass_2018,ohayon_precision_2022,delaunay_towards_2021}, and stringent tests of charge-parity-time (CPT) symmetry in anti-hydrogen \cite{bluhm_mathitcpt_1999,atrap_collaboration_first_2004,andresen_trapped_2010,atrap_collaboration_trapped_2012,kostelecky_lorentz_2015,malbrunot_asacusa_2018,ahmadi_characterization_2018,the_alpha_collaboration_investigation_2020}.  
\backmatter

\bmhead{Acknowledgments} 
The authors gratefully acknowledge funding of this project through NSF grant \#1654425. We also acknowledge useful conversations with Jacob Roberts and Samuel Brewer, and Mingzhong Wu for the loan of RF test equipment.


\bibliography{refs.bib}


\section{Methods}
\label{sec:model}
\subsection{Model}
For the Monte Carlo simulations, we calculate the AC Stark shifts by diagonalizing the interaction Hamiltonian. For these calculations, we include the 2S F=1 levels and all fine and hyperfine levels in the 4P manifold. We include all fine structure splittings but ignore the 4P hyperfine splittings. The 2S F=0 level is not considered in our analysis due to the selective two-photon excitation to the 2S  F=1 states.  The resulting AC Stark shift of the 2S level is an analytic function of the laser intensity and detuning which we use to calculate a time and spatially varying potential. We use this potential to simulate classical atomic trajectories. 


We model the lattice intensity as 
\begin{equation}
    I(\vec{r},t)=I_0(1+\eta^2+2\eta\cos{(\Phi(z,t))})e^{-(r/w_0)^2} e^{-((t-t_0)/t_w)^6},
    \label{eq:standingwave}
\end{equation}
where $I_0$ is the peak laser intensity, $\Phi$ is the axial phase of the standing wave interference pattern, $w_0$ is the Gaussian beam width and $\eta^2$ is the ratio of the backward to forward propagating laser power.  In general, $\eta \neq 1$ because the double-passed AOM and EOM each have a non-negligible optical loss. Since the lattice is pulsed, we include a super-gaussian temporal function centered at $t_0$ and having a duration of $t_w$. We neglect variation of the beam width in the z direction since the Rayleigh range of our interaction region is longer than the interaction region.  
The axial phase in Eqn. \ref{eq:standingwave} is given by
\begin{equation}
    \Phi(z,t)=2kz-\omega_{\Delta} t+\Phi_{c}\cos(2 \pi f_{c}t+\phi),    \label{eq:latticePhase}
\end{equation} 
where $\omega_\Delta$ is the frequency difference produced by the acousto-optic modulator, and $\Phi_c$ and $f_c$ are the respective modulation amplitude and frequency produced by the electro-optic modulator. The position of the lattice amplitude pulse with respect to the modulation is controlled by the phase offset $\phi$. The speed of the lattice is given by
\begin{equation}
     v^{(L)}(t)=v_{l}-v_{c}\sin(2 \pi f_{c}t+\phi),
    \label{eq:convenientPhaseVelocity}
\end{equation}
where we have introduced the central lattice velocity $v_l=\omega_\Delta/(2 k)$, and the amplitude of velocity modulation $v_c=\pi f_c \Phi_c/k$.  The lattice acceleration is given by
\begin{equation}
     a^{(L)}(t)=-a_0\cos(2 \pi f_{c}t+\phi),
    \label{eq:convenientPhaseVelocity}
\end{equation}
where $a_0=2 \pi f_{c} v_{c}$ is the acceleration amplitude. 
Experimentally, values for $v_c$, $a_0$ and $v_l$ were measured \emph{in situ} using a Michelson interferometer with a fast photodiode and oscilloscope. We numerically integrated classical particle trajectories using these parameters for the optical lattice. To account for spontaneous emission loss to the ground state we also numerically integrated the steady-state optical Bloch equations (OBE) for the population of 4P$_{1/2}$ and 4P$_{3/2}$ levels over the atomic trajectory through the optical lattice. Numerical simulations of millions of trajectories and survival probabilities are calculated over initial positions and velocities defined by our atomic beam \cite{cooper_cryogenic_2020}.

\subsection{Experiment}
The generation and characterization of our cryogenic beam of 2S hydrogen is described in previous work \cite{cooper_cryogenic_2020,cooper_cavity-enhanced_2018,burkley_highly_2019,brandt_measurement_2022}. The slowing laser at \SI{486}{\nm} has not been previously described and begins with an extended cavity diode laser (ECDL) at \SI{972}{\nano\meter} in the Littrow configuration. The output is amplified with a commercial tapered amplifier (DILAS), and frequency doubled to \SI{486}{\nano\meter} using an AR coated lithium triborate (LBO) crystal in a resonant bowtie cavity.  The \SI{486}{\nano\meter} source produced $\approx$\SI{970}{\milli\watt} of CW power.  After the optical losses from beam shaping and the Pockels cell, approximately $750-$\SI{850}{\milli\watt} power was available for the decelerator. We measured \SI{600}{\milli\watt} of backward going power due to losses in the double-pass AOM and EOM. The radiation is focused to an \SI{80}{\micro\meter} spot size at the atomic interaction region.  These values lead to $I_0 \approx 3.7$ kW/cm$^2$ and $\eta^2=0.8$.  


Amplitude control of the slowing laser was implemented using a Con-Optics model 380 EOM in an amplitude modulation configuration, which was driven using a home built MOSFET circuit. The system was capable of generating the $\approx$\SI{40}{\ns} pulses shown in Fig. \ref{fig:MCphase}. The MOSFET gate was controlled by a pulse generator with a maximum repetition frequency of \SI{1}{\mega\hertz}. Both the lattice chirp signal and the pulse generator were controlled by the same function generator so that the relative phase between the two signals could be phase-locked, adjusted, and recorded during data collection. 

To apply the frequency chirp and accelerate the lattice, the beam is double passed through a $3$ mm $\times$ $3$ mm $\times$ \SI{20}{\milli\meter} long y-cut phase modulation EOM made from magnesium doped lithium niobate with $r_{33} =$ \SI{32}{\pico\meter\per\volt}. 
The EOM is operated as a pure phase modulator and is resonantly driven by a step-up transformer at $6$ \SI{}{\mega\hertz}. The measured lattice acceleration data is shown in Fig. \ref{fig:MCphase}. 

For detection of the metastable atoms, a CEM detector is installed on-axis to provide a normalization signal.  The off-axis detector is placed off-axis by \SI{4}{\degree} a distance of \SI{27}{\centi\meter} from the slowing region. 


\end{document}